\documentclass[10pt,preprint2]{aastex}
\accepted{30 July 2001}
\shorttitle{The Radio Emission from Young Supernovae}
\shortauthors{Mioduszewski, Dwarkadas \& Ball }
\slugcomment{accepted for publication in ApJ}
\def\etal{et~al.}
\def\SSA{synchrotron self-absorption}
\def\FFA{free-free absorption}
\begin{document}

\title{Simulated Radio Images and Light Curves of Young Supernovae}
\author{Amy J. Mioduszewski\altaffilmark{1}, Vikram V. 
Dwarkadas\altaffilmark{2} \& Lewis Ball\altaffilmark{3}}
\affil{RCfTA, School of Physics, University of Sydney, NSW 2006, Australia}
\affil{~\\~\\ {\rm Accepted for publication in ApJ}}

\altaffiltext{1}{current address: National Radio Astronomy Observatory,
P.O. Box 0, Socorro, NM 87801, U.S.A.; e-mail address: amiodusz@nrao.edu}
\altaffiltext{2}{current address: Bartol Research Institute, Univ of Delaware,
217 Sharp Lab, Newark, DE 19716, U.S.A.; e-mail address: 
vikram@bartol.udel.edu}
\altaffiltext{3}{current address: Australia Telescope National 
Facility, CSIRO, PO Box 276, Parkes, NSW 2870, Australia; e-mail address: ball@physics.usyd.edu.au}
\begin{abstract}
We present calculations of the radio emission from supernovae based on
high-resolution simulations of the hydrodynamics and radiation
transfer, using simple energy density relations which link the
properties of the radiating electrons and the magnetic field to the
hydrodynamics.  As a specific example we model the emission from
SN1993J, which cannot be adequately fitted with the often-used
analytic mini-shell model, and present a good fit to the radio
evolution at a single frequency.  Both
\FFA\ and \SSA\ are needed to fit the light curve at early times and a
circumstellar density profile of $\rho \sim r ^{-1.7}$ provides the
best fit to the later data.  We show that the interaction of density
structures in the ejecta with the reverse supernova shock may produce
features in the radio light curves such as have been observed.  We
discuss the use of high-resolution radio images of supernovae to
distinguish between different absorption mechanisms and determine the
origin of specific light curve features.  Comparisons of VLBI images
of SN1993J with synthetic model images suggest that internal free-free
absorption completely obscures emission at 8.4~GHz passing through the
center of the supernova for the first few tens of years after
explosion.  We predict that at 8.4~GHz the internal free-free
absorption is currently declining, and that over the next $\sim 40$
years the surface brightness of the center of the source should
increase relative to the bright ring of emission seen in VLBI images.
Similar absorption in a nearby supernova would make the detection of a
radio pulsar at 1~GHz impossible for $\sim 150$ years after explosion.

\end{abstract}


\section{Introduction}

Although supernovae (SNe) have been observed over many centuries, it
is only in the past few decades that it has been possible to detect
them at radio wavelengths soon after outburst. Since the first radio
SN (RSN) detection by \citet{got72}, many more RSNe have been
observed, and a few have been discovered by virtue of their radio
emission alone (e.g.~SN1981K and SN1986J).  One of the earliest RSNe
to be discovered, SN1979C, has been extensively monitored at radio
wavelengths for almost two decades.  Many others, such as SN1986J,
SN1980K, SN1993J and the well-known SN1987A, have also been the
subject of extended studies.

\citet{wei86} and \citet{wei96} identified several common
characteristics of most RSNe:
a) the emission is non-thermal with very high brightness temperatures;
b) they appear first at high frequencies and then progressively at
lower frequencies;
c) at each frequency there is an initial rapid increase and then,
after reaching a maximum, the radio flux density declines as a
power-law in time;
d) the spectral index $\alpha$
(where the flux density $S\propto\nu^{-\alpha}$) between any
two frequencies is initially negative,
increases rapidly, and ultimately approaches a constant positive value.
It is generally accepted that the radio emission from SNe is synchrotron
radiation from ultra-relativistic electrons accelerated at the expanding
supernova shock (or shocks).
The declining power-law frequency spectrum seen after the SN radio emission
peaks is consistent with optically-thin synchrotron emission from a
power-law energy distribution of ultrarelativistic electrons.
The observed characteristics (a)--(d) are thus often loosely summarized by
the statement that radio SNe exhibit an optically-thick rise followed by an
optically-thin power-law decline.
Recent observations of deviations of RSNe light curves from long-term trends
have been attributed to changes in the ambient density
\citep{mon98,mon00,pan99}.

The radio emission from supernovae is thought to arise in the
hydrodynamically unstable interaction region between the SN ejecta and
a circumstellar medium (CSM) formed by mass loss from the progenitor
star.  This region contains both shocked supernova ejecta and shocked
swept-up material, separated by a contact discontinuity, and is
bounded on the outside by a forward shock and on the inside by a
reverse shock that travels back into the ejecta with respect to the
contact discontinuity.  The recent observation of a hard X-ray tail in
the spectrum of SN1006 has shown that electrons can be accelerated up
to TeV energies at a supernova shock \citep{koy95}.  The electrons may
be subject to a magnetic field which is essentially that of the
progenitor star wind, or which is amplified as a result of turbulence
behind the supernova shock.  However, there are as yet no detailed
models that satisfactorily treat the hydrodynamics of the supernova
expansion, the electron acceleration, and the origin and evolution of
the magnetic field, that together conspire to give the observed radio
properties.

An analytic treatment of the evolution of an idealized RSN
by \citet{che82a,che84} is often referred to as the ``mini-shell model''.
This model assumes that the expanding supernova is self-similar,
with the shock radius expanding as a power-law in time,
$r_s\propto t^m$, where $m$ is a constant \citep{che82a}.
The energy distribution of the relativistic electrons is assumed to be
a power law of constant spectral index,
and the energy density of the relativistic electrons and that of the
magnetic field are assumed to each be a constant fraction of the thermal
energy density behind the expanding supernova shock.
Free-free absorption in the ionized CSM \citep{che82b},
which decreases as the SN shock expands through the overlying CSM,
can produce an initial optically-thick increase in flux density.
Decreasing synchrotron self-absorption (SSA) in the expanding radio
source may also be important \citep{che95, che98}.

The mini-shell model, and a range of empirical variants of it
(cf. Van Dyk et al.~1994a; Weiler et al.~1996),
have been used to parameterize the radio
emission from a number of young supernovae.
The mini-shell model works well within its limitations \citep{che86}
and can successfully reproduce the evolution of some RSNe.
However, some of the empirical extensions of the mini-shell model
which have been used to fit recent RSNe no longer reflect the
physical basis of the original model.
Furthermore, the mini-shell model assumes self-similarity and is
therefore not applicable to RSNe whose ejecta are not represented by a
power-law density profile,
or whose expansion is not self-similar for other reasons.

Attempts to compute RSN models that incorporate a more detailed
treatment of the physical processes involved have generally focused
on trying to describe in more detail the acceleration of electrons
to relativistic energies.
\citet{duf95} considered a time-dependent treatment of
diffusive shock acceleration in their study of SN1987A.
They used a 2-fluid model in which the accelerated ions provided
an additional component to the total pressure.
\citet{fra98} adopted a
different approach in their study of SN1993J, opting to fit the
individual radio spectra at the various observed dates,
rather than concentrating on the light curves at different wavelengths.
Their calculated light curves are consistent with observations up to
an age of about 4 years.
\citet{jun99} have included relativistic electron transport
in their magnetohydrodynamic simulations of a young supernova remnant 
interacting
with an interstellar cloud.

We have developed a robust and general method to relate the
hydrodynamics of a supernova explosion to the evolving radio emission
from these sources.
We first compute the evolution of the SNe using spherically-symmetric,
high-resolution numerical hydrodynamic simulations.
Our method allows for the use of arbitrary profiles for the SN ejecta and
circumstellar medium, a significant generalization from analytic
models which are restricted to the self-similar expansion which results
from power-law ejecta and CSM profiles.
The light curves are produced by a
detailed radiative transfer calculation using ray tracing methods.
Synchrotron self-absorption is implicitly included in the calculation,
free-free absorption in the surrounding medium is explicitly
calculated, and in one case we include internal \FFA.
The relativistic particle energy density and the
magnetic energy density can be independently specified,
and in general are a function of radius.
In the simplest implementation, used for the results presented in this paper,
these are taken to be a fixed fraction of the local thermal energy density,
similar to the assumptions used in the mini-shell model.
A further generalization of the method, computing the local relativistic
particle distribution using advanced particle acceleration codes,
is already underway.

The results of calculations for a model for SN1980K with power-law
ejecta and CSM density profiles -- giving a self-similar expansion --
are presented as a test of our method.
We then model the emission from a model of SN1993J with a complicated ejecta
profile which is motivated by detailed modeling of the optical and x-ray
emission, and which results in an expansion that is far from self similar.


In this paper we first describe our method, give the results
of calculations of the radio emission from model RSNe,
and compare the model emission with the observational data
for SN1980K and SN1993J.
The radiation transfer calculation and the manner of computation are
outlined in \S \ref{radtr}. 
In \S \ref{comlc} we describe the process of calculating the model radio 
emission,
including the hydrodynamic simulations that are used as
an input to the radiative transfer code.
As a test of the method and the code we calculate the radio evolution
from a hydrodynamical simulation of a self-similar shock expansion,
and compare the results with the behavior expected from an analytic treatment
of this special case.
In \S \ref{fitlc} we discuss the process of fitting observed radio data,
and present the results of model calculations for SN1980K,
which can be fitted using the minishell model, and SN1993J which cannot.
Finally, in \S \ref{dis} we evaluate the results,
discuss the implications of this work,
and outline future prospects.

\section{Radiation Transfer} \label{radtr}

The basic radiation transfer equation (e.g. Rybicki \& Lightman 1979) is
\begin{equation}
{{\rm d}I_\nu(x) \over {\rm d}x} = j_\nu(x) - \kappa_\nu(x) I_\nu(x) , 
\label{transfer}
\end{equation}
where $I_\nu(x)$ is the flux density, $j_\nu(x)$ is the emissivity and
$\kappa_\nu(x)$ is the opacity, all at position $x$.
The optical depth between position $x_0$ and $x$ is defined as
\begin{equation}
 \tau_\nu(x_0,x) = \int_{x_0}^{x} \kappa_\nu(x^\prime) {\rm d}x^\prime.
\label{tau} 
\end{equation}
The solution to Equation (\ref{transfer}) between $x_0$ and $x$ can then be
written as
\begin{equation}
 I_\nu(x)=I_0(x_0) e^{-\tau_\nu(x_0,x)}
    +  \int_{x_0}^{x} e^{-\tau_\nu(x^\prime,x)} j_\nu(x^\prime) {\rm 
d}x^\prime.
\label{intensity1}
\end{equation}
We assume that there is no background radiation (i.e., $I_0 = 0$),
and denote the diameter of the SN by $r_s$ so that the equation we use
for our radiation transfer calculations is:
\begin{equation}
 I_\nu(x>r_s \sin\theta)=\int_0^{r_s\sin\theta}
  e^{-\tau_\nu(x^\prime,r_s\sin\theta)}
j_\nu(x^\prime) {\rm d}x^\prime
\label{intensity}
\end{equation}
where $\theta$ is the angle between the line of sight and the radial line
to the intersection of the line of sight and the shock.

The hydrodynamical simulations discussed here are 1-dimensional. The
code converts these data into a 3-dimensional sphere embedded in a
Cartesian grid.  The hydrodynamical quantities at each grid point are
used to calculate $\kappa_\nu$ and $j_\nu$ (as described in the next
two sections).  These are then integrated along each line of sight to
calculate $\tau_\nu$ and finally the flux density, $I_\nu(r)$ for that
line of sight.  The integration is performed using the trapezoidal
rule.  The numerical grid is sufficiently fine that the method is
adequate, and the trapezoidal rule is very robust.  A 2-dimensional
array of surface brightness (as a function of position across the
source) is built, and this array can be used to produce a synthetic
image or summed to obtain a point on a light curve.

\subsection{Synchrotron Emission \&\ Absorption}

The synchrotron emissivity ($j_\nu$) and opacity ($\kappa_{\nu,SSA}$)
depend on the distribution of the radiating electrons, the magnetic field 
($B$),
and the frequency ($\nu$). 
We assume that the energy distribution of ultra-relativistic electrons
is a power law $n(E)=n_0 E^{-\gamma}$ where $E$ is the particle energy,
$\gamma$ and $n_0$ are constants, and the total number density is
$\int n(E) {\rm d}E$.
%
%
%
The hydrodynamical simulations do not provide direct information 
on the magnetic field or the high energy particle distribution.
Here we follow \citet{che82b} and assume
that the magnetic energy density, $u_B$,
and the relativistic particle energy density,
$u_{rel}$,
are proportional to the thermal particle internal energy density,
$u_e$, such that
\begin{eqnarray}
u_B =& B^2/(2\mu_0) =\zeta_Bu_e,&
\label{zetaB}\\
u_{rel} =&\int n(E) \, E{\rm d} E =\zeta_{rel}u_e,
\label{zetarel}
\end{eqnarray}
where $\zeta_B$ and $\zeta_{rel}$ are 
constants.  In the \citet{che82b} mini-shell model these constants
are set to around 1\%.  The common assumption of energy equipartition
corresponds to $\zeta_B=\zeta_{rel}$.  There is some justification
for such a relation between the energy densities: $u_e$ is high between the
forward and reverse shock, which is where the magnetic field
is likely enhanced  and the particles accelerated to relativistic speeds.
However such arguments are weak and ultimately need to be replaced by
physical processes, such as particle acceleration.

The thermal
particle energy density is related to the thermal 
pressure ($p$), which is output by the hydrodynamic simulations,
through the adiabatic index ($\Gamma$):
\begin{equation}
p(r,t)=(\Gamma-1)u_e(r,t)
\label{pressure}
\end{equation}
Following \citet{mio97} equations (\ref{zetaB}), (\ref{zetarel}) and
(\ref{pressure})
can be used to write the synchrotron emissivity and opacity as
\begin{eqnarray}
j_\nu(r,t) \sim (\gamma-2) \Delta \,\zeta_B^{-{1\over 4}}\;
    \left({p(r,t) \over {\Gamma-1}}\right)^{{\gamma+5}\over 4} 
\nu^{-{\gamma-1\over 2}},\\
\kappa_{\nu,SSA}(r,t) \sim (\gamma-2) \Delta 
    \left({p(r,t) \over {\Gamma-1}}\right)^{{\gamma +6}\over 4} 
\nu^{-{\gamma+4 \over 2}},
\end{eqnarray}
where $\Gamma$, $\gamma$, $\zeta_B$, $\zeta_{rel}$ and 
$\nu$ are
input parameters,
$\Delta= \zeta_B^{{\gamma+2}\over 4} \zeta_{rel}$,
 and $p(r,t)$ is obtained from the hydrodynamic simulations.

\subsection{Free-Free Absorption}

In contrast to the synchrotron emission, free-free absorption (FFA)
depends
upon the thermal density and
temperature of the plasma and therefore can be calculated with a minimum of
assumptions.
The opacity ($\kappa_{\nu,FFA}$) goes as
\begin{equation}
\kappa_{\nu,FFA}\sim {n_e n_i \over n_\nu \nu^2 T^{3/2}} \;\ln C(T,\nu)
\label{kappaff}
\end{equation}
where $n_\nu$ is the refractive index and $C(T,\nu)$ is a constant 
\citep{lan80}.
For most of the models for RSNe considered here the \FFA\ arises from the
overlying circumstellar medium (CSM) and $n_\nu\sim 1$.
In the hydrodynamical simulations
the CSM density $\rho_{\rm CSM}$ is assumed to fall off as a power law, 
$\rho_{\rm CSM}\sim r^{-s}$ where $r$ is the radial distance from
the progenitor and $s$ is a constant.  We also assume that a constant 
fraction of 
$\rho_{\rm CSM}$ is ionized.
The temperature ($T$) of the
CSM is assumed to be constant with radius and is an input parameter.
The free-free optical depth,
$\tau_{\nu,FFA}$ is obtained by integrating from the outer edge of the SN,
determined by the radius of the forward shock ($r_s$), to infinity, whence:
\begin{equation}
\tau_{\nu,FFA}= {r_s^{1-2s}\over{2s-1}}\;K_{FFA}
\label{tauffa}
\end{equation}
where $K_{FFA}=\kappa_{\nu,FFA} r_s^{2s}$ is independent of $r_s$.

Although the external
temperature is a parameter in the hydrodynamical simulations, it does
not have a strong effect on the pressure between the forward and
reverse shocks (because the pressure of the CSM is much less than the 
pressure in the
post-shocked region), where most of the radiation is produced.
It is therefore reasonable to use a single hydrodynamical simulation to 
calculate the resulting synchrotron emission for a range of assumed CSM
temperatures.

For internal \FFA\ ccombined with \SSA\ (only considered in \S \ref{iffa}),
$\kappa_{\nu,FFA}$ is
added to $\kappa_{\nu,SSA}$ and the total $\kappa_{\nu}$ is used
in the radiation transfer calculation outlined at the beginning of \S 
\ref{radtr}.

\section{Computing Light Curves} \label{comlc}

We calculate light curves in two steps.  First, a hydrodynamical
simulation is run for a given SN ejecta profile and a given CSM profile.
The results of a simulation comprise the spatial profile of hydrodynamical
quantities at different epochs in the
evolution of the SN.
The particular outputs we require are the
pressure profiles at each epoch,
which are used as inputs to the radiation transfer code to calculate
the radio flux densities of the SN at each epoch.

\subsection{Hydrodynamic Simulations}

We model the hydrodynamics of the interaction of
the expanding supernova ejecta with the circumstellar medium
using high-resolution simulations.
All the simulations described here were carried out with the VH-1 code,
a 3D finite-difference code (here used as 1D)
based on the Piecewise Parabolic Method of
\citet{col84}.
The calculations are carried out in
Lagrangian co-ordinates and re-mapped onto an Eulerian grid.

The VH-1 code uses an expanding grid which is essential for simulations where
the dimensions change by orders of magnitude over a single run.
The grid tracks the outer shock and the outer grid radius increases 
accordingly.
The rate of expansion of the grid is adjusted such
that most of the grid is taken up by the interaction region, bounded
by the inner (or reverse) shock, and the outer (or forward) shock.
Resolving this high-pressure region in detail is the highest priority
since it is the source of most of the radio emission.
Over a large part of the evolution almost a half to two-thirds of the grid
is occupied by the interaction region, ensuring that it is
sufficiently well sampled.

Setting up the initial conditions for a simulation requires a model
for the SN ejecta and a model for the CSM that
the ejecta are to expand into.
For Type II supernovae the most commonly used ejecta models
have a density profile which is uniform inside some reference radius
and decreases as a power law with radius in the outer parts.
The interaction of power-law ejecta with a power-law circumstellar medium
can be described analytically by the self-similar solutions of
\citet{che82a} and \citet{nad85}.

If the CSM is formed via a constant wind from the progenitor star
before it exploded as a SN, then the wind density of this medium would
vary as $r^{-2}$ where $r$ is the radial distance from the star.
Ultimately, as the wind merges with the interstellar medium, the density
profile must approach a uniform value.
Many factors, such as evolution of the wind or structure in the pre-existing
ambient medium, can lead to different density distributions at smaller radii.

\subsection{Testing the code: Self-Similar Solutions}

We first test the code by computing the emission from a self-similarly
evolving SN which can be compared with analytic results. 
We consider SN ejecta with a power-law density profile
$\rho \propto v^{-n}$ where $\rho$ is the density, $v = r/t$ is the velocity
of the freely-expanding ejecta at radius $r$, and $n > 5$ is a constant.
When such ejecta expand into a
circumstellar medium with density ${\rho}_{CSM} \propto r^{-s}$, the
radius of the contact discontinuity ($r_{1}$) evolves as
$r_{1} \propto t^{(n-3)/(n-s)} = t^m$ \citep{che82a}.
The intrinsic (unabsorbed) flux density,
given the assumed scalings of the magnetic field and relativistic electrons,
then scales with frequency and time as
$S\propto \nu^{-\alpha} t^\beta$ where $\beta=-\alpha-3(1-m)$
\citep{che82b} and $\alpha=(\gamma-1)/2$.

\begin{deluxetable}{lcc}
\tablewidth{0pt}
\tablecolumns{3}
\tablecaption{Time Dependence Indices}
\tablehead{
  \multicolumn{1}{c}{} &
  \colhead{analytic} &
  \colhead{measured}}
\startdata
  no absorption & $-$1.05 &  $-$1.05 \\
  synchrotron-self absorption & 2.30 & 2.32   \\
  free-free absorption & $-$2.70 & $-$2.71  \\
\enddata
\tablecomments{The index quoted for the \FFA\ case is for
$\ln \left({F_{\rm thin}\over {F}}\right)$,
rather than for the flux density itself.
  }
\end{deluxetable}

Most observed SN light curves are characterized by a rapid,
optically-thick rise and a more gradual optically-thin decline (for
examples see \citet{wei96}).  The optically-thick rise in these models
is caused by decreasing absorption as the SN expands.  This absorption
can be due to \FFA\ and/or \SSA .  Since the \FFA\ is directly related
to the density of the overlying CSM, the decreasing CSM density with
radius causes the \FFA\ to decline.  Less directly, the \SSA\
decreases because the declining density in the CSM causes the the
pressure between the forward and reverse shocks to decrease and thus
the magnetic field and density of relativistic particles decrease.  We
consider three limits in which the time dependence of the radio flux
density can be expressed analytically
\citep{che82a,che84}:
\begin{enumerate}
\item optically thick due to internal \SSA
\begin{equation}
F \propto t^{(2-{2-s\over4})m + {1 \over 2}}
\label{thickSSA}
\end{equation}

\item optically thin (i.e.~with negligible optical depth due to either \SSA\ 
or \FFA)

\begin{equation}
F \propto t^{3m + [(2-s)m-2] {\alpha+3\over 2}}
\label{thin}
\end{equation}

\item optically thick due to external \FFA.

\begin{equation}
\ln \left({F_{\rm thin}\over {F}}\right) \propto t^{(1-2s)m}
\\
\label{thickFFA}
\end{equation}
where $F_{\rm thin}$ is the flux density in the absence of absorption.
\end{enumerate}
For the specific self-similar case we consider, $n=12$
and $s=2$ and we set $\gamma=2.5$. Thus $m=(n- 3)/(n-s)=0.9$.
Table 1 shows the analytic values of the time indices for these
three cases together with the values obtained from fitting the
light curves produced by the radiation transfer code for the
corresponding hydrodynamical simulation.
The agreement is remarkably good.

%
\begin{figure*}[!hbt]
\epsscale{2.0}
\plottwo{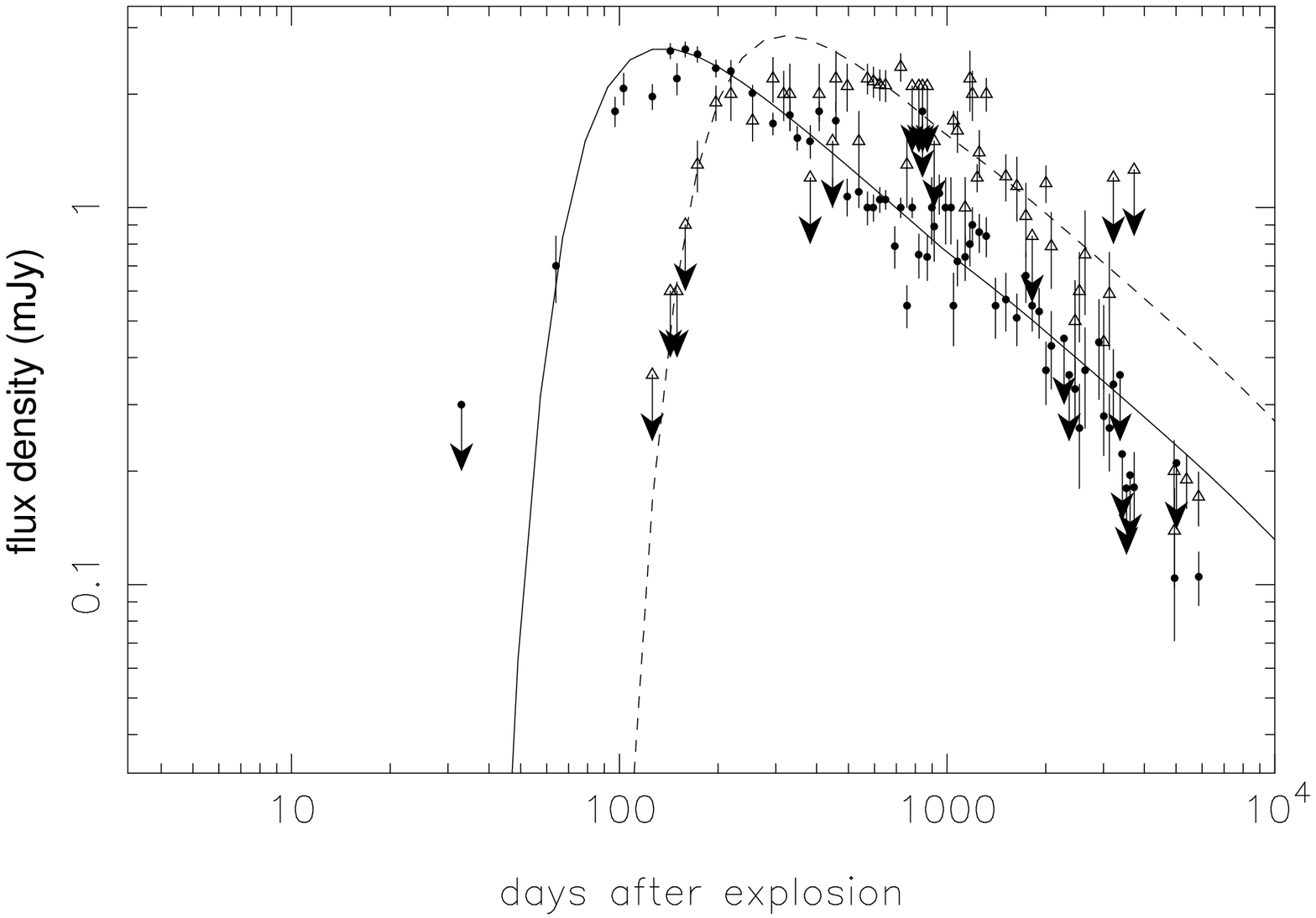}{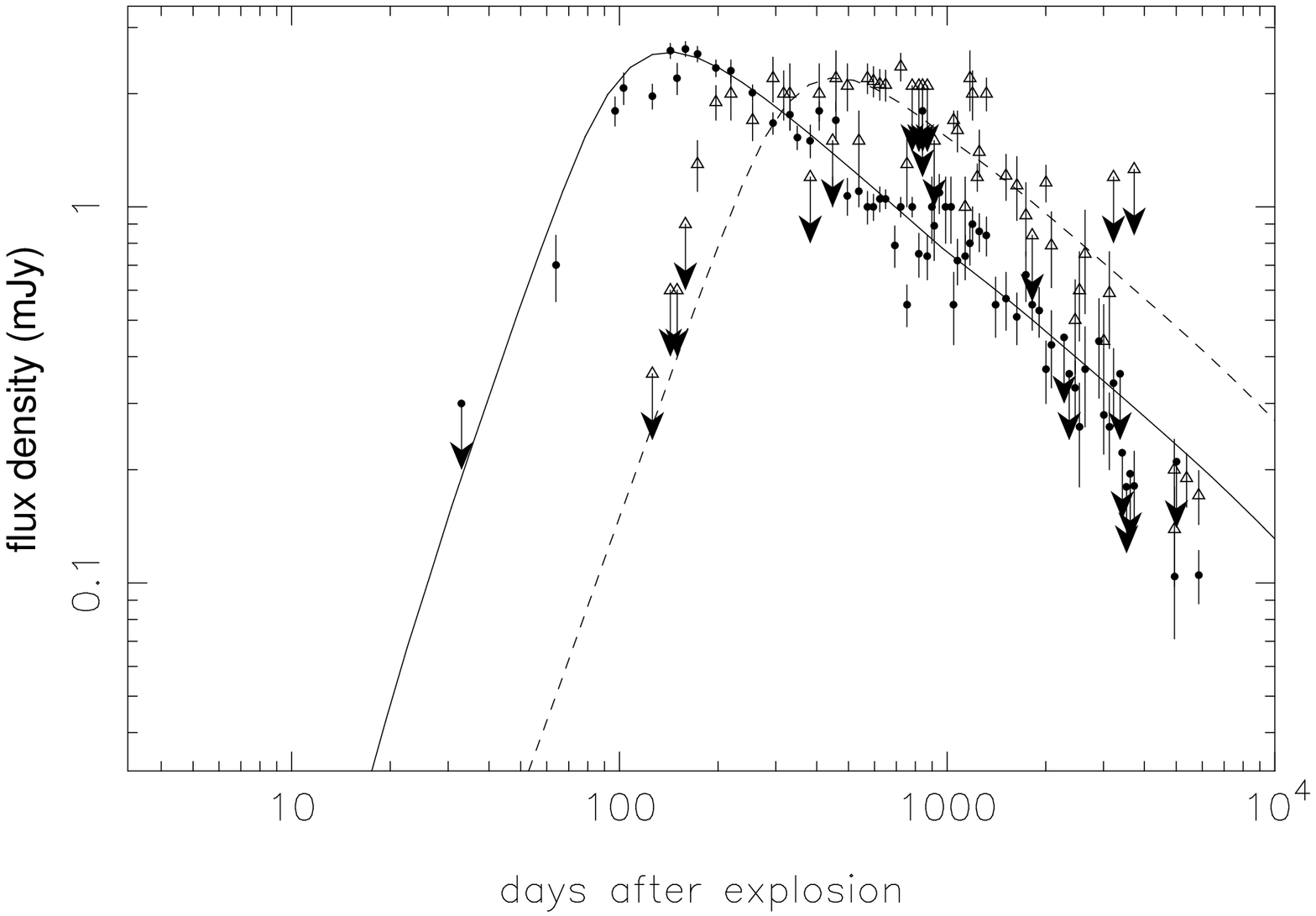}
\caption{Observed radio flux density of SN1980K (points)
and model flux densities (lines)
for the expansion of a power-law ejecta model
into an $r^{-2}$ CSM density profile and an $r^{-20}$ ejecta density
profile.
The 5 GHz data are shown as filled circles and models are shown as solid lines,
the 1.4 GHz data are open triangles and models are dashed lines.
The left panel shows a model with only free-free absorption in the CSM,
with $\chi^2=2.03$ and parameters $\alpha=0.6$ and
$T_{CSM}=3\times10^5\rm K$.
The right panel shows a model with only \SSA,
with $\chi^2=2.07$ and parameters $\alpha=0.6$ and
$\Delta=4.2\times 10^{-5}$.
}
\label{80K}
\epsscale{1.0}
\end{figure*}

\section{Fitting Light Curves} \label{fitlc}

Given the results of a
numerical hydrodynamical simulation, 4 further parameters must be
specified to calculate the evolving radio emission:  the spectral
index ($\alpha$ or $\gamma$), the temperature of the CSM ($T_{CSM}$), and
the energy density constants, $\zeta_B$ and $\zeta_{rel}$.
The different effects that these 4 parameters have on the light
curves, and their implications for fits to observed data,
can be summarized as follows:

\begin{description}

\item[$T_{CSM}$] influences only the external \FFA.
Most importantly, this parameter determines the time when the optical
depth is small enough that the source can first be observed
(i.e.\ when the SN turns on in the radio) if \FFA\ is the dominant
absorption mechanism at that time. 
Increasing the ionization fraction of the CSM
can mimic decreasing the temperature (see equation \ref{kappaff}).
A model in which the CSM temperature decreases with radius
could also produce a low optical depth early in the SN evolution.
The fitted model parameter $T_{CSM}$ therefore does not necessarily
reflect the true temperature of the CSM.

\item[$\gamma$ {\rm (or equivalently,} $\alpha${\rm )}]
determines the frequency dependence and the slope
of the optically-thin decline, as can be seen from inspection
of equation (\ref{thin}).  Its other
effects can be balanced by the constants, $\zeta_B$
and $\zeta_{rel}$.

\item[$\zeta_B$ {\rm and} $\zeta_{rel}$] affect the relative levels of the
optically-thin and thick parts of the light curve, but do not
change the slope of the rise or fall of the light curve. 
When the late optically-thick part of the light curve is dominated
by \SSA\ these parameters determine the position of the peak of the light 
curve.

We calculate the flux density in arbitrary units, and finally scale the light
curves to the observed data.
The values of $\zeta_B$ and $\zeta_{rel}$ must satisfy the physical restriction
$\zeta_{rel}+\zeta_B \leq 1$; but cannot be determined independently.
When \SSA\ is unimportant $\zeta_B$ and $\zeta_{rel}$ can be absorbed into the
overall scaling parameter and so are not determined by the model fit,
and are not quoted.
When \SSA\ is important, any two models with the the same value of
$\Delta=\zeta_{rel}\zeta_{B}^{\gamma+2\over 4}$
produce the same radio emission apart from an arbitrary scale factor.

\end{description}

We follow an iterative procedure to fit the observed evolution of
a radio supernova.
Having chosen and run an appropriate hydrodynamical simulation,
we choose a set of the four
additional model parameters and calculate the radiative transfer to
obtain the radio evolution.
The four parameters are adjusted on the basis of a comparison of the computed
radio emission with the observed data, and the radiative transfer calculation
is repeated for the same hydrodynamical simulation results.

\subsection{SN1980K} \label{sn80k}

SN1980K is a Type II supernova that was searched for at radio wavelengths
soon after optical maximum, but not detected until some 35 days later 
\citep{wei86}.
\citet{wei92} fit a Chevalier mini-shell model,
with free-free absorption in an $r^{-2}$ CSM,
to the data and obtain a reduced $\chi^2$ of 2.7.
A fit to a more extensive data set by \citet{mon98}
provides a similar value of reduced $\chi^2$,
but the model used is not directly
comparable to our calculations.

Our method is a generalization of the Chevalier mini-shell model,
and it provides a similarly satisfactory fit to the data
for a simple power-law ejecta density profile
expanding into a CSM with $\rho_{CSM}\propto r^{-2}$.
Figure \ref{80K} shows the observed radio evolution of
SN1980K at two frequencies together with two such model fits.
The left panel shows a model with only
free-free absorption in the CSM.
The implied $T_{CSM}$ is quite high ($3\times10^5\rm K$),
but is consistent with the value $T_{CSM}\sim 10^5-10^6\rm K$
found by \citet{lun88} at early times.
In any event, $T_{CSM}$ may not indicate the true temperature of the CSM,
as explained at the beginning of this section.
A fit using the same hydrodynamical model
when synchrotron self absorption is important, but there is 
negligible free-free absorption in the CSM,
is shown in the right panel of Figure \ref{80K}.
Qualitatively, the two alternative model fits to the 5\,GHz light curves
are very similar; the \FFA\ model fits the optically thick rise
at 1.4\,GHz better, but the \SSA\ model better matches the peak.
Quantitatively, the fits give comparable values of $\chi^2$
and we cannot distinguish between the two absorption mechanisms.
The $\chi^2$ values of both of these fits (calculated using the data
up to day 3380) are very similar to that of \citet{wei92}.

%
\begin{figure*}[!hbt]
\epsscale{2.0}
\plotone{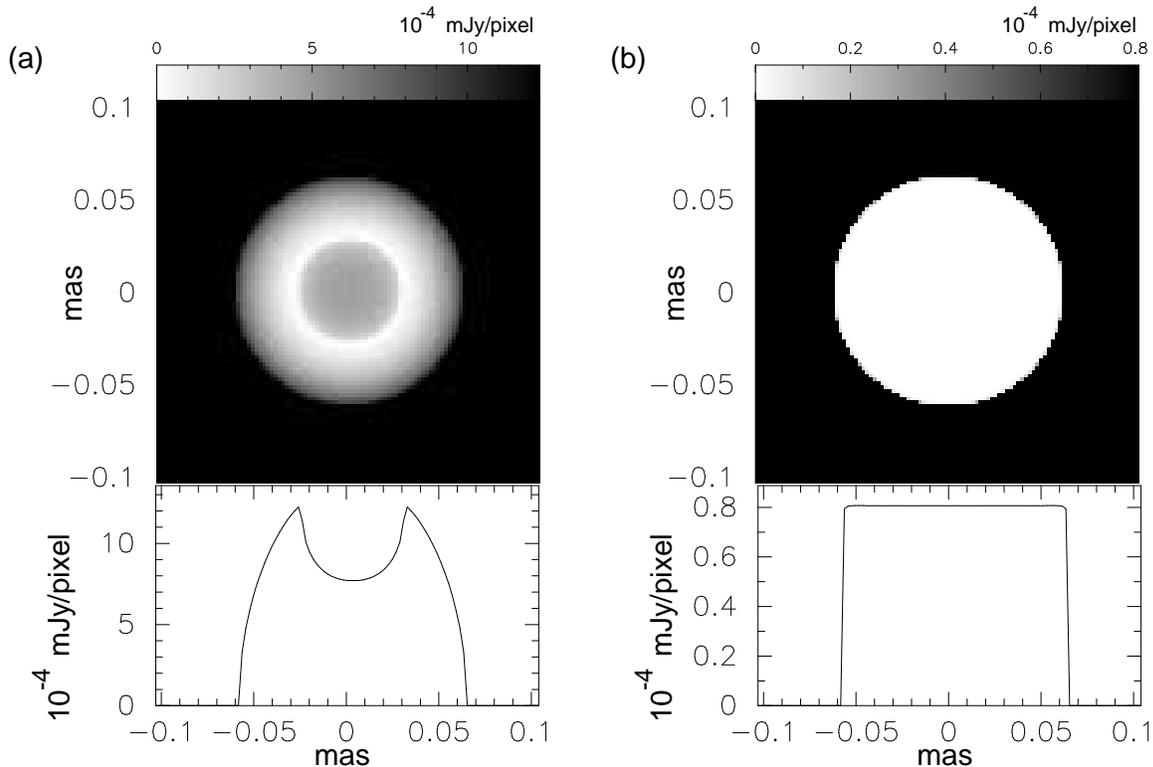}
\caption{Simulated images and surface brightness profiles of SN1980K
at 1.48 GHz on day 172.  The panel on the left shows the case when
only \FFA\ is important, and the panel on the right shows the case when
only \SSA\ is important.
}
\label{80Kim}
\epsscale{1.0}
\end{figure*}

\citet{che98} estimated the minimum radius of the SN at the
peak of the light curve if \SSA\ was the dominant absorption mechanism,
and argued \SSA\ was probably not significant in SN1980K
because the implied expansion velocity is much smaller than observed.
However, this argument assumes equipartition of energy between
the relativistic particle energy density and the magnetic
energy density (i.e., $\zeta_{rel}=\zeta_B$).
The model that we fit allows $\zeta_B$ and $\zeta_{rel}$ to take values
which are far from equipartition, in which case the arguments
of \citet{che98} imply an expansion velocity for SN1980K which is similar to
the measured velocity and \SSA\ is not ruled out.

Although neither the fits to the light curves,
nor the expansion velocity arguments, determine whether the observed
optically-thick rise of SN1980K is dominated by \FFA\ or
by \SSA, the spatial distribution of surface brightness in these
two cases is quite different.
Our method of calculation of the radio emission allows us to
generate synthetic images,
and two examples are shown in Figure \ref{80Kim}.
These synthetic radio images and the corresponding radial profiles of
the surface brightness shown below them are for the same simulations
used to produce the light curves in Figure \ref{80K}.
They show the emission at 1.48~GHz on day 172 after explosion,
when the source is still optically thick at this frequency.
The \FFA\ dominated image is a distinct ring which is brightest at a
radius roughly half the total radius of the source.
In contrast, the \SSA\ dominated image is a uniformly-bright disk.
These differences were predicted analytically by \citet{mar85}
and would be easily distinguishable (in principle)
via high resolution images at epochs very soon after the
explosion of a radio supernova.
Unfortunately, SN1980K could not have been imaged at sufficient
resolution to reveal these differences,
and even the two closest supernovae detected to date,
SN1993J and SN1987A,
were not imaged with sufficient resolution at
early times to make this distinction.

\subsection{SN1993J}

SN1993J, in the nearby galaxy M81 at a distance of 3.6 Mpc
\citep{fre94}, is one of
the closest SNe to explode in modern history.
It is believed to be a Type II SN that
may have been a member of a binary system
\citep{hof93, pod93, woo94}.
The radio evolution  of SN1993J has been followed in detail
throughout most of its lifetime, and thanks to its proximity it has
been possible to image the evolving RSN extensively with VLBI
\citep{mar97, bar00}.
\citet{mar97} found that the expansion of the radio source
was consistent with a self-similar $r\propto t^m$ dependence.
However, \citet{bar00}, using careful astrometry on a larger sample of
observations obtained over a longer time period,
have reported that the expansion is not self-similar.
The simple mini-shell model cannot reproduce the rapid turn-on and
subsequent relatively slow optically-thick rise of the radio emission
from SN1993J \citep{vdy94b}.
Furthermore, since the evolution of the SN1993J radio source
is not self-similar,
a direct application of the mini-shell model is not appropriate.
It is therefore an ideal test case for our method of calculation of the
radio emission.

%
%
\begin{figure*}[ht!!]
\epsscale{2.0}
\plotone{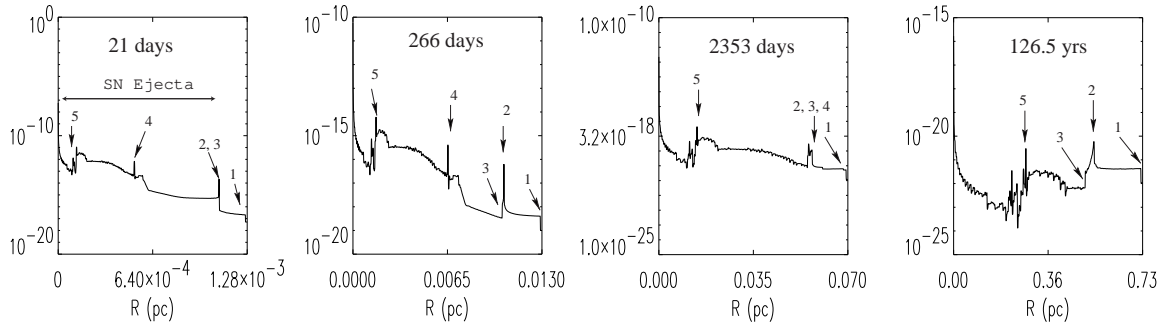}
\caption{Density profiles at four epochs from the hydrodynamic simulation
of the 4H47 ejecta model for SN1993J expanding into a circumstellar medium with
$\rho_{CSM}\propto r^{-1.7}$. The numbered labels indicate the following 
features:
1.\ forward supernova shock; 2.\ contact discontinuity; 3.\ reverse shock;
4.\ outer local ejecta maximum; 5.\ inner local ejecta maximum.
The collision of the local maximum (4) with the reverse shock is shown in the 
third panel.}
\label{93Jprofiles}
\epsscale{1.0}
\end{figure*}

\citet{fra98} assumed a power-law ejecta density profile
for SN1993J and produced simulated radio light curves which compare well
with the early data.
However, Nomoto, Suzuki, Iwamoto and other coworkers have suggested that
the optical and X-ray evolution of SN1993J implies that the ejecta
distribution is more complicated.
\citet{iwa97} computed the 2D hydrodynamical evolution of
SN1993J and calculated the optical light curve.
They found a series of ejecta models that gave an acceptable fit to the 
observed
light curve if the extent of the mixing for nickel was assumed to be larger 
than
seen in their 2D simulations.
Similar models were used by \citet{suz95} to fit the
X-ray emission with reasonable results.

We use a specific ejecta model supplied by this group
-- model 4H47 (courtesy of Ken Nomoto, private communication) --
and calculate the hydrodynamics and
radio emission from the model supernova.
The model ejecta mass is 3.12 M$_{\odot}$,
the pre-SN progenitor star has a radius of 350 R$_{\odot}$
and an envelope mass of 0.47 M$_{\odot}$.
The fraction of helium in the envelope mass was about 0.79.
The explosion energy is 10$^{51}$ ergs.

The density distribution of the CSM surrounding SN1993J is
poorly constrained by observations.
Many of the models that reproduce the optical, X-ray and radio
emission require a density profile that falls off more slowly than
$r^{-2}$ (eg. Suzuki \& Nomoto
1995; Fransson, Lundqvist \& Chevalier 1996; Nomoto and Suzuki 1998;
Van Dyk et al.~1998).
\citet{fra98} however find the
standard $r^{-2}$ distribution to be adequate in their self similar
modeling of the radio light curve.
We therefore consider three CSM distributions,
with density varying as $r^{-1.5}, r^{-1.7}$, and $r^{-2}$.

Figure \ref{93Jprofiles} shows the density profiles at four timesteps during 
the
hydrodynamic evolution for a CSM density profile of $\rho_{CSM}\propto 
r^{-1.7}$.
The interaction of the ejecta with the CSM leads to the formation of an
outer shock which expands through the CSM,
and a reverse shock that propagates back into the ejecta,
separated by a contact discontinuity.
The 4H47 ejecta model contains local maxima,
two of which are indicated by arrows in the top left panel of
Figure \ref{93Jprofiles}.
The inclusion of a cooling function in our
hydrodynamical simulation means that these peaks remain well defined
as the ejecta expand.
There is a significant density jump just ahead of the outermost
local density maximum (labelled 4). The interactions of this
density jump and the local peak with the reverse shock occur
at ages of around 1300 days and 2300 days respectively,
the latter of which is close to the present epoch.
These interactions have a dramatic effect on the pressure profiles,
and hence on the expansion of the supernova and the radio emission.

The structure in the ejecta profile leads to a distinctly non-self-similar
evolution.
We define the expansion parameter $M(t)={\rm d}\ln r_s/{\rm d} \ln t$ so that 
for
self-similar expansion $M(t)=m$, a constant.
Figure \ref{expansion} shows the variation of the expansion parameter
with time during the early stages of the
simulation of the interaction of the 4H47 ejecta with
the $r^{-1.7}$ CSM.
The decrease of the expansion parameter with time over the first 2300 days,
representing an increasing rate of deceleration of the supernova shock,
is consistent with the observations of \citet{bar00} who fit an
expansion parameter $M(t)\sim~ 0.937$ up to 306 days
and $M(t)\sim 0.77$ between 582 and 1893 days (shown as dashed
lines in figure \ref{expansion}).
At around day 2300 the interaction of the outermost local maximum in the ejecta
density profile with the reverse shock leads to a significant increase in
the pressure between the reverse and forward shocks.
This has the effect of accelerating the forward shock
and leads to the rapid increase in the expansion parameter seen in
Figure \ref{expansion} around days 2300--4000.

\subsubsection{Radio Supernova 1993J}

The radio evolution of SN1993J is most extensively 
followed at 8.4~GHz.
We investigate the radio emission calculated from the hydrodynamical simulation
of the 4H47 ejecta model by first trying to fit the observed 8.4~GHz
light curve, and then considering the data at other frequencies.

%
%
\begin{figure}[!h]
\plotone{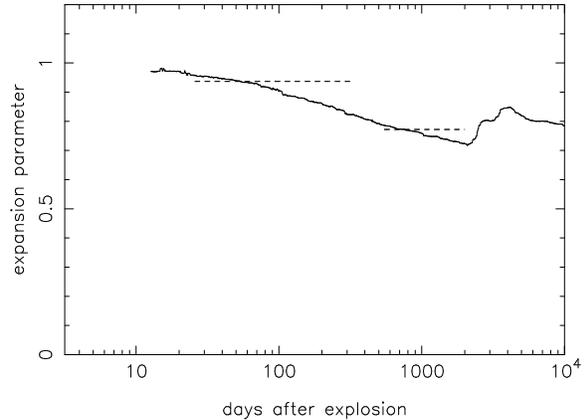}
\caption{Evolution of the expansion parameter $M(t)$ (solid line)
calculated from the
hydrodynamical simulation of the 4H47 model ejecta expanding into a
CSM with a density profile falling off as $r^{-1.7}$.  The horizontal
dashed lines are the expansion parameters fit by \citep{bar00} in
the time ranges they are valid.
}
\label{expansion}
\end{figure}

\begin{figure}[!hbt]
\plotone{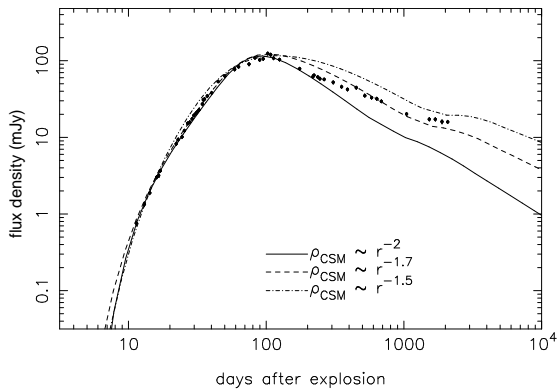}
\caption{Observed radio flux density of SN1993J at 8.4~GHz (filled circles)
with model fits (lines) using both \FFA\ and \SSA\
from three simulations of the expansion of the 4H47 model ejecta
into different CSM density profiles as labelled.
The model parameters are:
solid line $\rho_{CSM}\propto r^{-2}$, $T_{CSM}=5\times 10^5 \rm K$,
$\Delta=3.27\times 10^{-4}$
and $\gamma=2.1$;
dashed line $\rho_{CSM}\propto r^{-1.7}$, $T_{CSM}=7\times 10^4 \rm K$,
$\Delta=6.66\times 10^{-4}$
and $\gamma=2.1$;
dash-dotted line $\rho_{CSM}\propto r^{-1.5}$, $T_{CSM}=1.2 \times 10^3 \rm K$,
$\Delta=1.06\times 10^{-3}$
and $\gamma=2.1$}
\label{bothabs}
\end{figure}

None of the fits obtained using only \FFA, or only \SSA, provides a
good match to the early data for any of the CSM profiles considered.
A much better fit is possible when both \FFA\ and \SSA\ are included.
Figure \ref{bothabs} shows the best fits obtained for
$\rho_{CSM}\propto r^{-2}$ (solid line), $\rho_{CSM}\propto r^{-1.7}$
(dashed line), and $\rho_{CSM}\propto r^{-1.5}$ (dash-dotted line).
All provide an adequate fit to the rising phase, an indication of the
flexibility provided by the two different absorption processes.  The
differences between the models become apparent near the peak of the
light curves, and increase as the flux density decreases.  In the
declining phase neither absorption process is important, and the only
model parameters that affect the light curve for a given
hydrodynamical simulation are the spectral index $\gamma$ (or
$\alpha$), $\Delta$ and the overall scaling.  The fit for a CSM
density profile of $\rho_{CSM}\propto r^{-2}$ is very good at times up
to about day 200, soon after the peak flux density, but at later times
the model flux density drops off more rapidly than the observations.
The CSM density profile of $\rho_{CSM}\propto r^{-1.7}$ provides the
best fit up to the present epoch.  The decline of the model for
$\rho_{CSM}\propto r^{-1.5}$ is considerably slower than the observed
dependence, and the peak is much broader than is observed.

%

For our best fit model, $T_{CSM}=7\times 10^4 \rm\ K$,
which is smaller than the value $2 \times 10^5 \rm K$ quoted by
\citet{fra98}, and than the values suggested by \citet{fra96}
who considered the CSM temperature in detail.
For $T_{CSM}< 7\times 10^4 \rm\ K$ the free-free opacity in our model
is too high at early times. 

All the model curves exhibit a temporary flattening in the decline,
at times varying from day $\sim1000$ for $\rho_{CSM}\propto r^{-2}$,
to day $\sim 2000$ for $\rho_{CSM}\propto r^{-1.5}$.
This is directly attributable to the interaction of the outer density
jump and local maximum in the 4H47 model ejecta with the reverse supernova 
shock,
which temporarily halts the decline in the pressure between the
forward and reverse supernova shocks.
The observed 8.4~GHz data show a similar flattening,
starting around day 1500 and extending to at least day 2500.
Such features in observed RSN light curves have previously been
attributed to interactions between the SN shock with density
structures in the CSM.
Our simulations suggest that light curve features may also arise from
interactions between the SN shock and density structures in the ejecta,
but it will be difficult to distinguish between these two possibilities
solely on the basis of the light curves.

%
\begin{figure*}[!hbt]
\epsscale{2.0}
\plotone{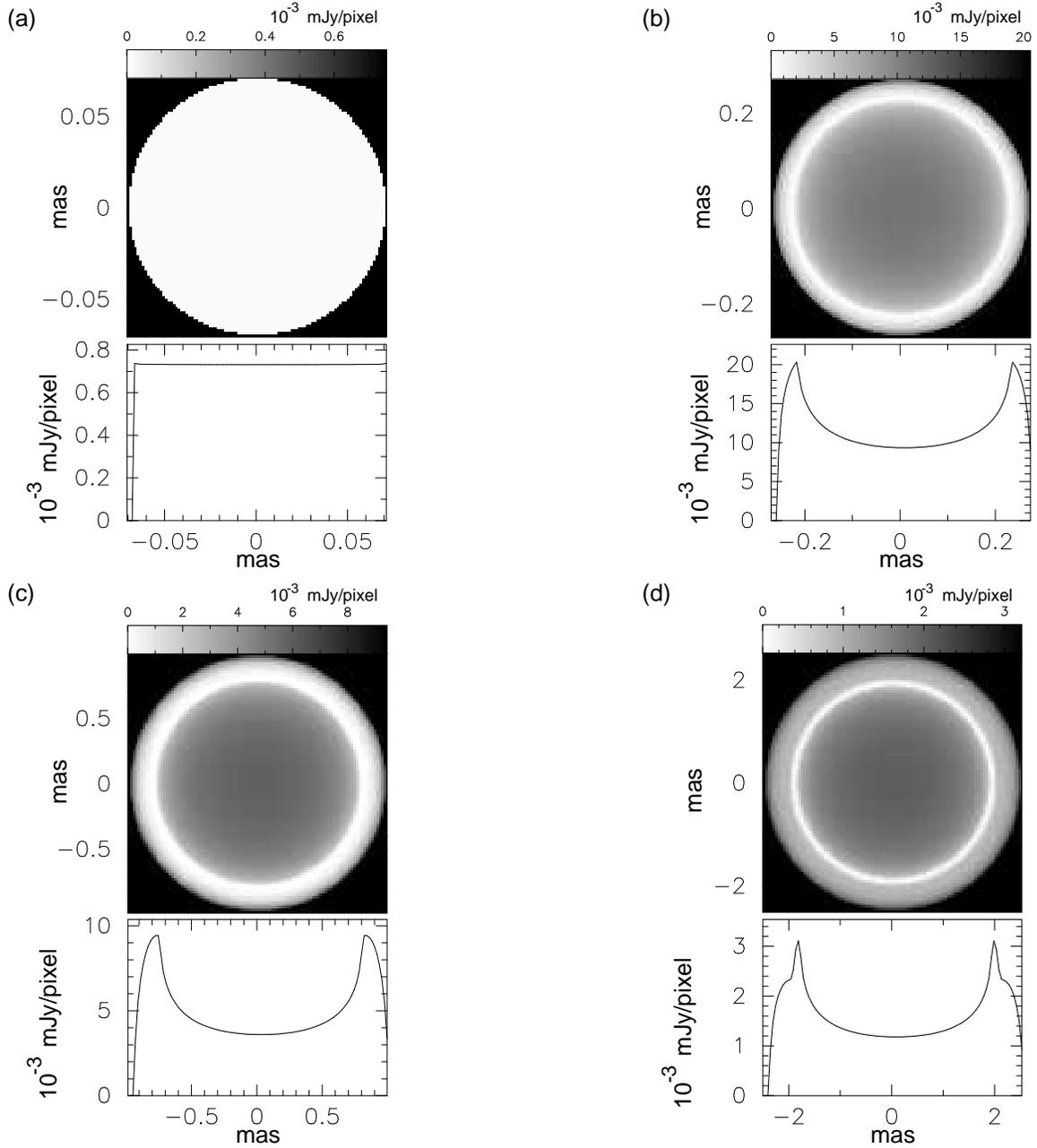}
\caption{Simulated images and surface brightness profiles of SN1993J
at 8.4~GHz for the same $\rho_{CSM}\sim r^{-1.7}$
simulation shown in Figure \ref{bothabs}.
The panels show the emission at different epochs as follows:
(a) day 20; (b) day 86; (c) day 390; and (d) day 1301.
}
\label{93Jim}
\epsscale{1.0}
\end{figure*}
           
Figure \ref{93Jim} shows synthetic images and surface brightness profiles of
SN1993J at 8.4~GHz at 20, 86, 390 and 1301 days after explosion,
for the the same $\rho_{CSM}\propto r^{-1.7}$
simulation as for Figures \ref{expansion} and \ref{bothabs}.
On day 20 -- panel (a) -- the source is optically thick and
\SSA\ dominates.   As expected, this
image is similar in character to the right panel in Figure \ref{80Kim}. 
On day 86 -- panel (b) -- SN1993J is becoming optically thin.
On day 390 -- panel (c) -- the source and the CSM are optically thin
and the flux density at all frequencies is declining with time.
On day 1301 -- panel (d) -- the model 8.4~GHz light curve has flattened
as a result of the interaction of the density jump in the ejecta with
the reverse shock.  As expected, the ring seen in 
Figure \ref{93Jim} a, b and c is bound on the inside by the reverse shock,
and on the outside by the forward shock.
The VLBI images of SN1993J of \citet{bar00} reveal a clumpy ring
of emission that when averaged azimuthally is qualitatively similar
to the simulated images shown in Figure \ref{93Jim}.
The radius and thickness of the bright ring  of the synthetic images are
similar to those of the observed data, but the radial profile of
the surface brightness in the synthetic images is quantitatively different
from that observed, a point we discuss in more detail in \S \ref{iffa}.
Panel (d) shows a distinct brightening on the inner edge of the
ring as a result of the pressure increase associated with the
interaction of a density discontinuity in the ejecta with the reverse SN shock,
which also causes the flattening seen in the model light curve.
If the observed flattening were due to an interaction between the forward
SN shock and a density structure in the CSM, the result would be a
brightening on the outer edge of the emission ring.
Such differences may ultimately provide a means of distinguishing
between SN shock interactions with structures in the ejecta and
those in the CSM.
However images of SN1993J obtained to date are not of sufficiently
high resolution to show these kinds of differences,
and the clumpy nature of the images complicates their interpretation.

%
\begin{figure}[!hbt]
\plotone{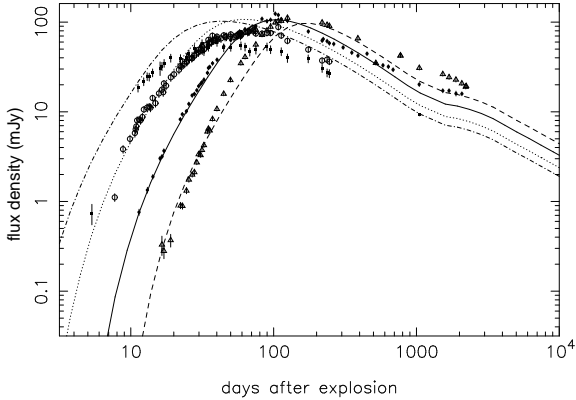}
\caption{Observed radio flux density of SN1993J (points)
and model flux densities (lines) for the expansion of the 4H47 ejecta model
into an $r^{-1.7}$ CSM density profile.
The different frequencies are indicated as follows:
22.5~GHz filled boxes and dot-dashed line, 15.0~GHz open circles and dotted 
line,
8.4~GHz filled circles and solid line, 4.9~GHz open triangles and dashed line.
The model parameters are as for the dashed line in Figure \ref{bothabs}.
}
\label{allfreq}
\end{figure}

Although Figure \ref{bothabs} shows the model flux densities for only a
single frequency, the model parameters determine the flux density at all 
frequencies
and at all epochs.
There are useful observations of the evolving radio flux density from SN1993J 
at
four frequencies, 4.9, 8.4, 15.0 and 22~GHz.
Figure \ref{allfreq} shows the observed data and the model flux densities
for the parameters determined by the best fit in Figure \ref{bothabs}
with $\rho_{CSM}\propto r^{-1.7}$.
The limitations of the model are immediately apparent.
Despite the relatively good fit of the model to the data at 8.4~GHz,
the fits at the other frequencies are relatively poor at all times
after the flux density has increased to a significant fraction of the maximum.
This may be attributed to the model itself,
not the particular choice of parameters used.
The fundamental model assumptions that the
magnetic energy density is proportional to the thermal energy density,
and that the relativistic electron energy distribution is a (fixed) power law
with an energy density which is also proportional to the thermal energy 
density,
provide an unbreakable link between the frequency and time dependence
of the model flux density.
At late times when the source and the CSM are optically thin
both the frequency and the time dependence of the model flux density
are completely determined by the model parameter $\alpha$ (or $\gamma$)
for a given hydrodynamical simulation.
The other model parameters, $T_{CSM}$, $\Delta$, and the overall scaling,
have no effect on the frequency or temporal dependence at late times.
In particular, the model fit to the 8.4~GHz data requires the choice
$\gamma=2.1$ to fit the observed temporal evolution.
This value of $\gamma$ implies $\alpha=0.55$,
which determines the frequency dependence of the model.
If the simulation was self similar the model flux density
in the optically-thin regime would therefore have
a frequency dependence of $S\propto \nu^{-0.55}$.
A simple fit to the observed spectra at times after
about 150 days indicates $S\propto \nu^{-0.8}$, a difference
which is much too large to be adequately fitted by the simulation.

%
\begin{figure}[!hbt]
\plotone{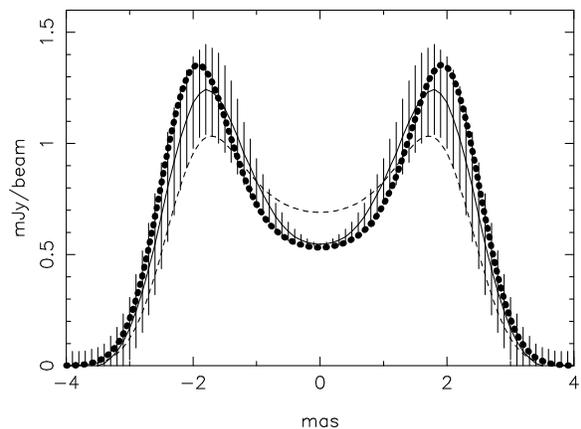}
\caption{The azimuthally averaged profile of the observed
8.4~GHz emission from SN1993J on day 1349 (solid line, Bartel et al.\
2000) with error bars showing the RMS scatter in azimuth, and the profile
of the simulation from day 1301 (convolved with the
same beam as the image) excluding (dashed line) and
including (dotted line) internal \FFA.
}
\label{profiles}
\end{figure}
\subsubsection{Internal Free-Free Absorption} \label{iffa}

\begin{figure}[!hbt]
\plotone{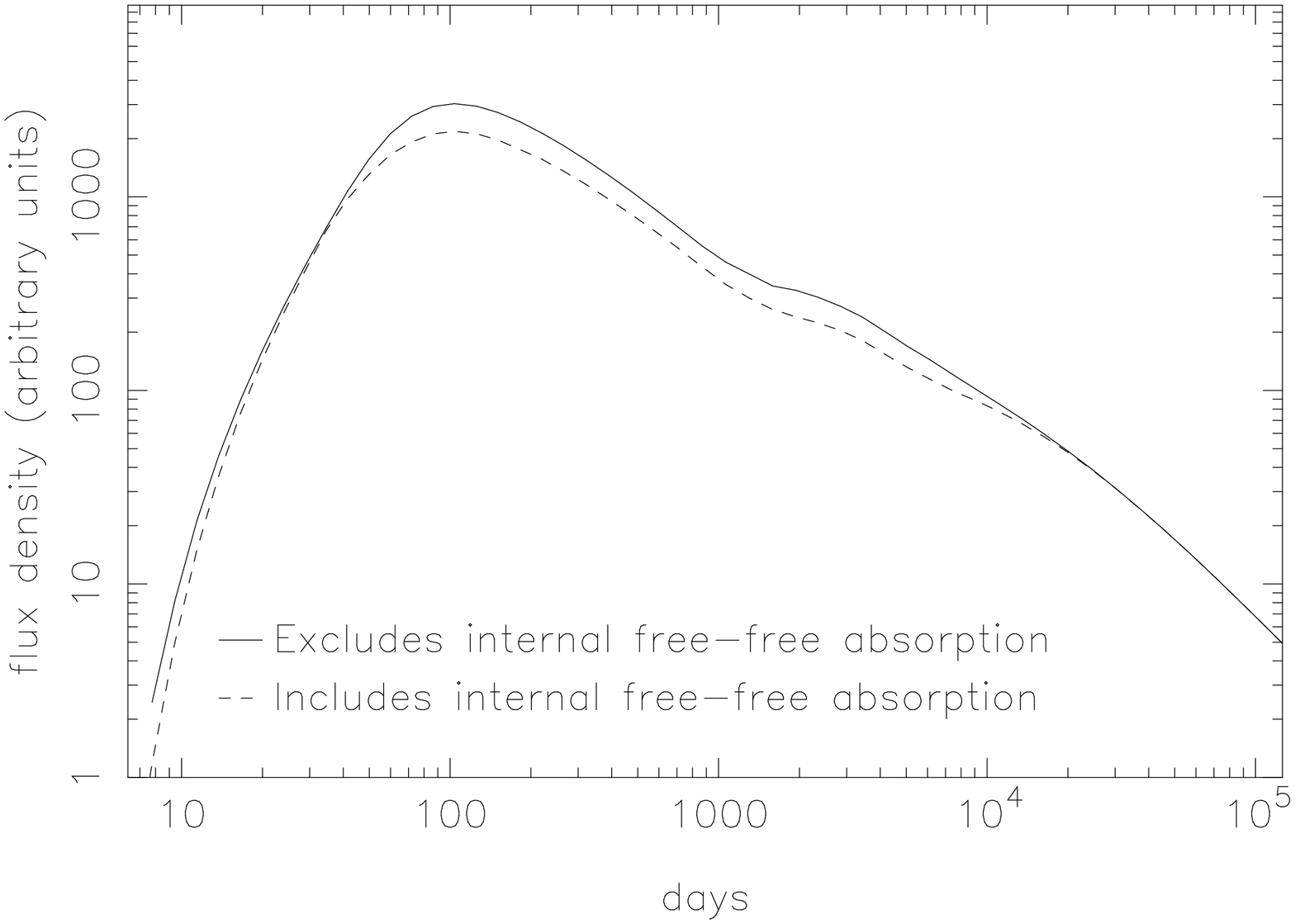}
\caption{Model 8.4~GHz light curves excluding (solid line)
and including (dashed line) internal \FFA\
for the same simulation used to generate the dashed line in
Figure \ref{bothabs}, and for Figures \ref{93Jim} \& \ref{allfreq}.
}
\label{iffalc}
\end{figure}

Figure \ref{profiles} shows the azimuthally averaged profile of 
the observed 8.4~GHz emission from SN1993J
(solid line) plotted with the profile of the
best fit model simulation at
approximately the same age (dashed line).  The diameter and
width of the bright ring of emission in the observed and
synthetic images are very similar.
However the ratio of the surface brightness at the center
of the image and at the brightest point on the ring is significantly
smaller in the observed data than in the simulated images.
The most likely reason for this difference is that internal \FFA,
i.e.\ free-free absorption occurring within the SN,
has not been included in our simulations.  
For the first tens of years the temperatures
and densities internal to the reverse shock are such that most,
if not all, the emission that passes through the interior is absorbed,
significantly decreasing the surface brightness seen from the center of the SN.
The dotted line in Figure \ref{profiles} shows the surface brightness profile
of the simulated image when internal \FFA\ is included.
It matches the observed profile well, lying within
the RMS scatter at all radii,
and we conclude that internal \FFA\ is important in SN1993J.
Figure \ref{iffalc} shows the comparison of the simulated 8.4~GHz
light curves for SN1993J with and without internal \FFA.
For the the first 8 or so years the light curves have a similar shape,
not changing our conclusions in the previous sections of this paper.
After year 8 there is a slight flattening in the light
curve until year 55 as the internal \FFA\ becomes unimportant due
to the decrease in the density inside the reverse shock as the SN
expands.
During this time (starting at around the present epoch) the ratio of
the surface brightness at the center of the SN to that at the peak
of the bright ring of the simulated SN increases steadily,
and we therefore
expect the same to occur in the VLBI images of SN1993J.
At higher frequencies the internal \FFA\ becomes unimportant at earlier 
epochs.
These conclusions and predictions regarding the internal \FFA\ are
independent of the assumptions used to calculate the synchrotron radiation.
If the surface brightness ratio does not increase as predicted, this has
implications for the internal density and temperature of the SN.

%

If other RSNe have a similar level of internal \FFA\ to that implied
for SN1993J, radio emission from an embedded pulsar would
be completely absorbed until the SN ejecta have expanded sufficiently
for the internal free-free optical depth to become negligible.
Our model for SN1993J suggests that the optical depth at 1~GHz falls
to unity around 150 years (50 years for 8.4~GHz) after the explosion.
Detection of a radio pulsar embedded in a similar nearby SN
would be impossible before that time.
For most extragalactic SN this is a moot point since radio pulsars
at extragalactic distances are unlikely to be detectable
even if not absorbed.
Modeling the radio plerion created by the pulsar, \citet{rey84} also
predict that one could not observe a pulsar for 200 years at radio frequencies
due to internal \FFA.

\section{Discussion and conclusions} \label{dis}

Our results indicate the viability of
the calculation of supernova radio emission
from high-resolution hydrodynamical simulations.
The calculations presented in this paper retain 
much of the essential physics of the mini-shell model,
particularly the connection
between the model parameters and physical parameters describing the
shock, the magnetic field, and the radiating electron distribution.
When the ejecta and the CSM have simple power-law density profiles
our generalized treatment reproduces the
results of the analytic self-similar model.
The advantage of our new method is that it provides a means to
calculate the radio emission in cases where the expansion
in not self-similar, which we demonstrate with calculations for
an ejecta model for SN1993J which has a complicated profile
motivated by detailed modeling of the optical and x-ray
emission.

Attempts to fit the observed time and frequency dependence
of the radio emission from SN1993J,
which is not well modeled by the analytic mini-shell,
are encouraging but not yet completely successful.
Fits to the evolving radio flux density of SN1993J
at a single frequency indicate that
neither \SSA\ nor \FFA\ alone can produce the observed
increase, but that when both effects are included
the rise in the flux density
can be well modeled for a range of parameters.
This conclusion reinforces the results of
\citet{fra98} and \citet{che98}
who argued
that both \SSA\ and \FFA\ were important for SN1993J.
Most other RSNe first become detectable at much later
epochs than did SN1993J, and their subsequent radio evolution
can be fitted adequately using only \FFA . 
This may be because in other SNe the conditions are such that
the optical depth due to \SSA\ is negligible by the time the
free-free optical depth through the CSM becomes small enough
for the SN to be observed.
An alternative is that the sparse sampling of the optically-thick
increase of most RSNe simply precludes a determination of the
relative importance of \SSA\ and \FFA ,
as shown in the SN1980K model fits of \S \ref{sn80k}.

Our simulations show that the different absorption mechanisms
produce quite different spatial surface brightness distributions
at very early epochs.
They also show that features in the radio light curves of SNe may
result from the interaction of density structures in the supernova
ejecta with the reverse supernova shock. Such interactions may be
detectable through visible signatures in the simulated images, such as
a brightening at one edge of the ring of radio emission.
VLBI imaging to date does not have sufficient resolution to reveal
such effects, but they may prove fruitful goals for future studies. 
However comparison of the profiles of VLBI images of SN1993J at 8.4~GHz
\citep{bar00} with our simulated images implies that internal
\FFA\ is important now and will remain so for a few tens
of years.
We  predict that the ratio of the surface brightness at the center of the SN
to that at the bright ring should increase,
starting at about the present epoch,
due to the gradual decrease of the internal \FFA\ as the SN expands.
A similar level of internal \FFA\ in a nearby SN would
absorb the emission from an embedded radio pulsar,
rendering it undetectable at 1~GHz until around 150 years after explosion.

The relatively slow decline of the SN1993J flux density
at a given frequency
may indicate that the magnetic field in the emitting
region declines more slowly than the scaling with thermal energy
density implies (e.g.\ Chevalier 1998),
or that the density of the CSM encountered by the expanding
supernova shock to date declines more slowly than the $r^{-2}$
dependence expected for a constant wind.
We find a good fit for the latter case with $\rho_{CSM}\propto r^{-1.7}$.
A density distribution less steep than
$r^{-2}$ has also been proposed by a number of other researchers
\citep{suz95, fra96, vdy98}.
Such a dependence could result from a time variation in the mass-loss
rate of the wind emitted by the pre-SN progenitor star, or a variation
in the wind velocity, or both. If the wind was steady during an
earlier phase of evolution of the progenitor, then at some distance
further away from the star one would expect the CSM density
distribution to revert to an $r^{-2}$ profile. 
\citet{suz95} and \citet{fra96} suggest that, in order to
match the X-ray and optical light curves and spectra, the density
distribution must change close to the star, within a radius of
about 10$^{16}$ cm. However, there is no break in the radio light
curves at an early stage that would indicate a sudden change in the
density distribution, and in our simulations we assume that the
initial CSM density distribution persists throughout
the evolution.

Finally, we find that the generalization of the mini-shell model
which results from a realistic treatment of the hydrodynamics
is insufficient to account for the observed evolution and
flux density dependence of SN1993J.
The link between the frequency and time dependencies of the
model flux density  due to the assumed energy density scalings
remains implicit in the implementation of the generalized model
used here, which therefore
doesn't have sufficient flexibility to fit the data.
The inclusion of any one of a number of aspects of the physics
which is not addressed by the model as presented here
may break this nexus.
An obvious next step is to investigate alternative scalings for the
magnetic field and/or the relativistic electron distribution
(see for example Chevalier 1996 and Chevalier 1998).
An alternative we are presently pursuing is to include
a physically-justifiable calculation of the time-dependent
acceleration of the radiating electrons resulting from the
calculated evolution of the expanding supernova shock,
along the lines of the treatment by \citet{duf95}.

\acknowledgments{
We thank the VLBI collaboration led by Norbert Bartel,
and in particular Michael Bietenholz and Michael Rupen,
for providing us with the flux density measurements of
SN1993J prior to publication and with access to their VLBI images,
and Ken-Ichi Nomoto for providing the 4H47 ejecta model.
AJM thanks Michael Rupen and Mark Wardle for suggestions which
greatly improved this paper.
VVD thanks Roger Chevalier for helpful discussions.
We are grateful to Kurt Weiler and colleagues for
providing RSN data sets on the web page
\url{http://rsd-www.nrl.navy.mil/7214/weiler/sne-home.html}.
This research has made use of NASA's Astrophysics Data System
Bibliographic Services. VVD's research was partially supported by a
grant from NASA administered by the American Astronomical Society.  }

\end{document}